\documentstyle[11pt,oldlfont]{article}

\def\1{\mbox{1\hspace{-.8ex}1}}

\def\g{\gamma}

\def\d{\delta}
\def\e{\epsilon}

\def\ve{\varepsilon}
\def\f{\varphi}

\def\D{\Delta}
\def\k{\kappa}

\def\lb{\label}

\def\m{\mu}
\def\n{\nu}
\def\c{\cite}

\def\pt{\cdot}
\def\om{\omega}
\def\r{\rho}

\def\Fb{\mbox{\boldmath$F$}}

\def\st{\mbox{\boldmath$*$}}
\def\dia{\mbox{\small \boldmath$\#$}}

\def\S{\mbox{\boldmath$\Sigma$}}

\def\X{\mbox{\boldmath$\Theta$}}

\def\Ab{\mbox{\boldmath$A$}}
\def\Cb{\mbox{\boldmath$C$}}

\def\eb{\mbox{\boldmath$e$}}
\def\ub{\mbox{\boldmath$\ub$}}

\def\fib{\mbox{\boldmath$\phi$}}
\def\t{\mbox{\boldmath$\tau$}}

\def\fb{\mbox{\boldmath$\f$}}

\def\Fib{\mbox{\boldmath$\Phi$}}
\def\ub{\mbox{\boldmath$u$}}

\def\w{\mbox{\boldmath$\omega$}}
\def\we{\mbox{ \boldmath$\wedge$}}
\def\L{\mbox{\boldmath$L$}}

\def\J{\mbox{\boldmath$J$}}
\def\U{\mbox{\boldmath$U$}}
\def\Bb{\mbox{\boldmath$B$}}
\def\Eb{\mbox{\boldmath$E$}}
\def\Gb{\mbox{\boldmath$G$}}
\def\Wb{\mbox{\boldmath$W$}}

\def\Db{\mbox{\boldmath$\Delta$}}
\def\sug{\mbox{\it \scriptsize sug}}
\def\bb{\mbox{\boldmath$b$}}
\def\ab{\mbox{\boldmath$a$}}
\def\beb{\mbox{\boldmath$\beta$}}
\def\alb{\mbox{\boldmath$\alpha$}}
\def\pb{\mbox{\boldmath$\psi$}}
\def\i{\mbox{\boldmath$i$}}
\def\R{\mbox{\boldmath$R$}}

\def\ex{\mbox{\boldmath$d$}}
\def\gab{\mbox{\boldmath$\g$}}
\def\cex{\mbox{\boldmath$D$}}
\def\Ob{\mbox{\boldmath$\Omega$}}

\def\bi{\bibitem}

\def\B{\begin{equation}}
\def\E{\end{equation}}

\begin{document}

\thispagestyle{empty}

\begin{flushright}   hep-th/9911035 \\
                    LPTENS 99/43 \\
                     AEI-1999-33  \end{flushright}

\vspace*{0.5cm}

\begin{center}{\LARGE {On first order formulations of supergravities}}

\vskip1cm
B. Julia$^a$  and S. Silva$^b$

\vskip0.5cm

$^a$Laboratoire de Physique Th{\'e}orique CNRS-ENS\\ 
24 rue Lhomond, F-75231 Paris Cedex 05, France\footnote{UMR 8549 du
CNRS et de l'{\'E}cole Normale Sup{\'e}rieure.
This work has been partly supported by the EU TMR contract
ERBFMRXCT96-0012.
}\\
\vskip0.2cm
$^b$Max Planck Institut f{\"u}r Gravitationsphysik, Albert Einstein Institut,\\
Am M{\"u}hlenberg 5, D-14476 Golm, Germany

\vskip0.5cm

\begin{minipage}{12cm}\footnotesize

{\bf ABSTRACT}

\bigskip

Supergravities are usually presented in a so-called 1.5 order
formulation. Here we present a general scheme to derive pure
$1^{st}$ order formulations of supergravities from the 1.5 order ones.  
The example of ${\cal{N}}_4=1$ supergravity will be rederived and
new results for  ${\cal{N}}_4=2$ and ${\cal{N}}_{11}=1$ will be
presented.

It seems that beyond four dimensions the auxiliary fields introduced to 
 obtain first order formulations of SUGRA theories do not admit
supergeometrical transformation laws at least before a full superfield
treatment. On the other hand first order formalisms simplify
eventually symmetry analysis and the study of dimensional reductions.
\bigskip
\end{minipage}
\end{center}
\newpage

\section{Introduction}
Since the discovery of the central importance of U-dualities to control
the divergences of string theories \cite{HTo,To,Wi} and of the duality
between Large N super-Yang-Mills theory  and AdS compactification of
eleven dimensional supergravity \cite{Ma} the need for a better
conceptual understanding of the latter has become rather    urgent.
The superspace approach is notoriously hard but component approaches
are
rather cumbersome, this is unsatisfactory as more miracles are  being
discovered \cite{CJLP2}. Finally the tensor calculus in 10 or 11
dimensions
is strongly restricted by supersymmetry so it is important to
streamline the corresponding constraints for instance on allowed
counterterms
and nonperturbative effects \cite{GV, DJST}.

The purpose of the present manuscript is to pursue the investigation
of the general structure of supergravities in their first order
formalism. In section 2, we will present a general method to
derive  first order formalisms. We will then treat the case of supergravities
in  general  in section 3. The exemples of ${\cal{N}}_4=1$,
${\cal{N}}_4=2$ and ${\cal{N}}_{11}=1$ will then illustrate the
procedure, sections 4, 5 and 6. 

A first order formalism for the three form of eleven dimensional
supergravity was recently
developped independently and with a different technique in \cite{NVN}, a first
order formalism for the Lorentz connection in eleven dimensions had
also been found in \cite{CFGPN}. However
our approach is both totally general and in the 11d case combines these
two results in a nicely symmetrical fashion.

\section{From $1.5$ to $1^{st}$ order formalisms}
Supergravity theories are conveniently written in a 1.5 order
formulation. It
simply means that the Lagrangian is a $1^{st}$ order Lagrangian, namely
the spin connection is to be varied as an independent field, but
on the other hand its local 
supersymmetry transformation law is not required and usually not known 
explicitly. 
The invariance of the action is only established after extremisation 
over the connection \cite{VN}.   
Both MacDowell and Mansouri \cite{DM} and 
Chamseddine and West \cite{CW} checked that the transformation
laws of Deser and
Zumino \cite{DZ} are geometrical up to the variation of the Lorentz connection,
they
went on to notice that using the latter's equation of motion it does not
matter, this remark applies equally well to the second order action of 
Freedman, van Nieuwenhuizen and Ferrara \cite{FNF}, this has been called 1.5 order
formalism see
for instance \cite{VN}. 
The group manifold approach in its original form or
in the modified form of \cite{DFTN} does not suffice to obtain our
complete results.

% \cite{CW}, \cite{TV}

The purpose of this section is to show a simple way to 
construct, from a 1.5 formulation, a $1^{st}$ order formulation, namely, to 
give a local supersymmetry transformation law for the spin connection,
that leaves the Lagrangian invariant.

Let $L (\f^a, \om^i)$ be a first order Lagrangian, which depends on some dynamical 
fields $\f^a$ and on  the auxiliary fields $\om^i$ (the spin connection in the 
case of (super)gravity). The latter fields can be eliminated algebraically
by making use of 
their equations of motion, that is:
\begin{equation}\label{rrr}
 E_{i} (\f,\om) := \left. \frac{\partial L}{\partial \om^i}\right|_{\f} = 0 \Rightarrow \om^i=\om^i_2 (\f)
\end{equation}
We used the fact that the Lagrangian can be written (up to a surface term)
as an analytic  function of $\om^{i}$ but not of its derivatives.

Due to the analyticity of the Lagrangian in the auxiliary fields, it is
possible to perform a Taylor expansion around $\om^i_2$. Defining $\D
\om^i (\f, \om) :=\om^i-\om_2^i (\f)$, we obtain\footnote{We will use the
Einstein summation convention for the $_{i}$ indices (and {\it not} the De
Witt one which includes  spacetime integration).}: 
\begin{equation}\label{taylor}
L (\f^a, \om^i) = L (\f^a, \om^i_2) +
% \D\om^i\left. \frac{\partialL}{\partial \om^i}\right|_{\om=\om_2}+
\frac{\D\om^i \D\om^j}{2} X_{ij}(\f^a, \om^i) 
\end{equation}
where 
\begin{equation}\label{defres}
X_{ij}(\f^a, \om^i):=\sum_{n=0}^{\infty}\frac{2}{(2+n)!}\left.\frac{\partial^{2+n}
L}{\partial \om^i \partial \om^j \partial \om^{k_{1}}\ldots \partial
\om^{k_{n}}}\right|_{\om=\om_2} \D \om^{k_{1}}\ldots\D \om^{k_{n}}
\end{equation}
 Note that the term linear in $\D\om^i$ in the Taylor expansion
(\ref{taylor}) vanishes identically due to (\ref{rrr}).

Let us
 suppose now that the action  $\int L (\f^a, \om^i_2)$ is invariant, as a 
$2^{nd}$ order functional, under a gauge symmetry, given by 
$\d_2 \f^a = \d_2 \f^a ( \f)$. Our purpose will now be to show
that it is possible to define a transformation law $\d_1$
that extends 
this gauge invariance to the $1^{st}$ order action $\d_1 \int L (\f^a,
\om^i)=0$. Let us first define $\d_1\f^a$  for simplicity by:

\begin{equation}\label{defd1}
 \d_1\f^a := \d_2 \f^a
\end{equation}

We need now a transformation law for the auxiliary fields
$\delta_{1}\om^{i}$ which, together with (\ref{defd1}), leaves the
$1^{st}$ order Lagrangian invariant. To proceed, let us vary the equation
(\ref{taylor}) with respect to $\d_1$. Using equation (\ref{defd1})
and the gauge invariance of the $2^{nd}$ order Lagrangian under
$\d_2$, we find that:

\begin{equation}\label{varlag2}
\d_1 L (\f,\om) = \d_1 \left( \frac{\D\om^i \D\om^j}{2} X_{ij}\right) =
\D\om^i X_{ij} \left(\d_1 \D\om^j  + \frac{1}{2}\D\om^l \bar{X}^{jk}\d_1 X_{kl} \right)
\end{equation}
Where $\bar{X}^{ij}$ is the inverse of the matrix $X_{ij}$.

Now, the $1^{st}$ order Lagrangian will be invariant under the
variation $\d_1$ if we impose the 
vanishing of (\ref{varlag2}) (up to some total derivative). This
condition is implemented if the transformation law of the auxiliary fields
is taken to be:

\begin{equation}\label{delome}
\d_1 \om^i = \d_2 \om^i_{2}  -  \frac{1}{2}\D\om^k \bar{X}^{ij}\d_1
X_{jk}
\end{equation}

This
formula shows that the ($1^{st}$ order) transformation law for the auxiliary
field $\omega^{i}$ is just the induced transformation law on
$\omega^{i}_{2}$, plus a term which vanishes modulo its equation of
motion. This general result was presented in \cite{H3}, without the
explicit formula (\ref{delome}). 

%This would be enough to define a $1^{st}$ order formulation from the 1.5 formulation. 

\bigskip 

In general, $X_{ij}$ depends on $\om^i$ and some algebraic
manipulations may be necessary to extract $\d_1 \om^i$ from equation
(\ref{delome}). In the simple cases where
the Lagrangian can be written (up to a
surface term) as a polynomial of degree 2 in the auxiliary
fields\footnote{This will be the case for all the applications treated here.},
i.e. $L=\frac{1}{2}\omega^{i}\omega^{j} X_{ij}(\f)-\omega^{i} X_{i} (\f) +X_{0}(\f)$
(with $X_{ij}(\f)$, $X_{i}(\f)$ and $X_{0}(\f)$ being arbitrary functionals of the
remaining dynamical fields and their derivatives), the above
equation (\ref{delome}) reduces with the help of (\ref{defd1}) to:

\begin{equation}\label{delome00}
\delta _{1} \omega^{i} = \delta _{2}
\omega^{i}_{2} (\f) + \frac{1}{2}  \delta _{2} \bar{X}^{ij} (\f) E_{j}
\end{equation}
 
This simple formula will be used in the following examples to find
$1^{st}$ order supergravities. It provides a first solution to our problem.

\bigskip 

Sometimes however it is advisable to rewrite the transformation laws by 
making use of a twist, that is, a trivial gauge transformation in the
terminology of \c{HT} (see also  \c{H3}). In fact  we can always redefine:
\begin{eqnarray}
\bar{\d}_1 \f^a &:=& \d_1 \f^a + \Omega^{ai} E_i (\f,\om) \label{twi1}\\
\bar{\d}_1 \om^i &:=&\d_1 \om^i - \Omega^{ai} E_a (\f,\om) \label{twi2}
\end{eqnarray}
or/and
\begin{equation}\label{entriv}
\bar{\d}_1 \om^i=\d_1 \om^i + \Xi^{ij} E_j
\end{equation}
and similarly for the $\f^{a}$'s.

Then, $\bar{\d}_1$ is still a gauge symmetry of the $1^{st}$ order
Lagrangian (for any $\Omega^{ai}$ and any antisymmetric matrix
$\Xi^{ij}=\Xi^{[ij]}$). Note that the trivial gauge transformations
(\ref{twi1}-\ref{entriv}) are just the simplest examples (which will
be enough for our present purpose). In fact,
these formulas remain unchanged in the De Witt notation, ie with a spacetime
integration in addition to the Einstein summation, see for example
\cite{HT}, \cite{SiT}.

This 
kind of twist will be used to simplify the form of  $\d_1 \f^a$. 
The examples of ${\cal{N}}_4 = 1$, ${\cal{N}}_4 = 2$ and ${\cal{N}}_{11} = 1$ 
supergravities are treated in the following sections. 
Note that due to the symmetry of the matrix $\bar{X}^{ij}$ (see the definition
(\ref{defres})), the last term of equation (\ref{delome00}) cannot be
twisted away.

\bigskip

We would like to conclude this section by the following observation:
associated to a $2^{nd}$ order formulation, there are an infinite
number of corresponding $1.5$ order theories. This is due to the
fact that the choice of the auxiliary field is somewhat
arbitrary. In fact, it is
straightforward to verify that a redefinition of the type
$\tilde{\om}^{i}:=\om^{i} + f^{i}(\f)$, where $f^{i}(\f)$ is an
arbitrary function of the fields $\f^{a}$ and eventually of their
derivatives, does not change the associated $2^{nd}$ order theory. In
the case of supergravities, there is a natural choice for the
auxiliary gravitational connection which is the supercovariant (``hatted'')
connection (see below the origin of supercovariance in 1.5 formalism). 
The same is true for the  supercovariant ``hatted'' field
strength, if any, as we shall see below.

\section{The case of supergravities}

The purpose of this section is to apply the previous formulas to derive
$1^{st}$ order formulations for {\it all} supergravities from their
$1.5$ formulations. We indeed give a general formula for the
supersymmetry variation of the connection for {\it any} supergravity
in {\it any} dimension. Explicit calculations will then be given for
the special cases of ${\cal{N}}_4=1$, ${\cal{N}}_4=2$ and
${\cal{N}}_{11}=1$ supergravities in the next sections. 

\subsection{Generalities on supergravities}\label{gengra}

The bosonic fields of supergravities are generically the one-form vierbein $\eb^a$
and eventually some (non)-Abelian $p$-forms $\Ab$'s. The auxiliary
fields associated to $\eb^a$ and to the $\Ab$'s are respectively  the
one-form spin-connection $\w^a_{\ b}$ and the $(p+1)$-form field
strength $\Fb$. Our aim is thus  to
compute the supersymmetry transformation laws for $\w^a_{\ b}$ and
$\Fb$ using formula (\ref{delome00}). This equation shows that 
the relevant part of the supergravity Lagrangian is the one quadratic in the
auxiliary fields. For {\it any} supergravity this is:

\begin{equation}\label{sugras}
\L_{\sug} (\fb,\w,\Fb) = -\frac{1}{4\k^2} \R^{ab} \we \S_{ab} -
\frac{1}{2} \Fb\wedge \st \Fb + \mbox{\it other} 
\end{equation}
Where $\R^{ab} := \ex \w^{ab} + \w^a_{\ c}\we \w^{cb}$ and 
$\S_{a_1 \ldots a_r} := \frac{1}{(D-r)!} \ve_{a_1 \ldots
a_ra_{r+1}\ldots a_D} \eb^{a_{r+1}}  \we\dots\we \eb^{a_D}$ (see also
\cite{JS} for notation and conventions). By {\it
other} we mean the rest of our supergravity action which is at most
linear in the spin connection and in the field strength. Here $\fb$
denotes all the remaining fields, i.e. those of the second order formalism.

As a consequence of equation (\ref{sugras}), the
Euler-Lagrange equations associated to the spin-connection can always
be written as:
\begin{equation}\label{eqw1}
\frac{\d \L_{\sug}}{\d \w^{ab}} = 
-\frac{1}{2\k^2}\D \w^c_{\ [a} \we \S_{b]c} = 0
\end{equation}
Where $\D \w^a_{\ b} =\w^a_{\ b}-\w^a_{2 b}$, $\w^a_{2 b} (\fb)$ being not
an independent field but the unique connection which satisfies the
equations of motion (\ref{eqw1}).

As the connection is an auxiliary field, the first equation
(\ref{eqw1}) can be algebraically inverted for $D \geq 3$. In components, we find that:

$$ \D \omega_{\m ab} = M_{\m ab}^{\n cd} \frac{\d L_{\sug}}{\d \omega^{\n cd}}  $$

Where,
\begin{equation}\label{defm}
M_{\m ab}^{\n cd} = 2 \k^2 \left( e_a^\n e_\m^{[c} \d_b^{d]} -
e^\n_b e_\m^{[c} \d_a^{d]}
-\d_\m^\n \d_b^{[c} \d_a^{d]} +\frac{2}{D-2} e^{\n [c} \left( \d^{d]}_a e_{\m b} -\d^{d]}_b e_{\m a} \right) \right)
\end{equation}

\subsection{The supersymmetry transformation law of the connection}\label{susytrc}

The $2^{nd}$ order supergravity Lagrangian $\L_{\sug} (\fb,\w_{2},\Fb_{2})$
is invariant under some local supersymmetry defined by its action on the
non-auxiliary fields namely $\d_{2}\fb$. We derive now the
supersymmetry transformation law of the spin-connection for the
$1^{st}$ order Lagrangian. The case of the field strength will be
treated in the next subsection.

 The method presented in the
previous section to compute $\delta _{1} \w^{a}_{\ b}$ from (\ref{delome00})
can be
summarized as the following recipe: ``Keep the part of the Lagrangian
quadratic in the connection (namely the
auxiliary field). Replace $\w^{a}_{\ b}$ by the difference $\D \w^{a}_{\ b}$
(defined after equation (\ref{eqw1})) in this piece of
the  Lagrangian. Then, impose the
vanishing of its variation under $\delta_{1}$''.
From (\ref{sugras}), the only part of the supergravity Lagrangian
which is proportional to the square of the connection $\w^{a}_{\ b}$
comes from the purely gravitational Einstein-Hilbert piece, namely
$\w^{a}_{\ c}\we \w^{cb}\we \S_{ab}$. Thus the previous recipe gives:
\begin{equation}\label{www}
\delta_{1} \left( \D \w^{a}_{\ c}\we \D \w^{cb}\we \S_{ab}\right)=0,
\end{equation}
and then, using assumption (\ref{defd1}),

\begin{equation}\label{varcon11}
\d_1 \w^c_{\ [a} \we \S_{b]c} = \d_2 \w^c_{2 [a} (\f)\we \S_{b]c} -\frac{1}{2}
\D \w^c_{\ [a} \we \d_2\S_{b]c}.
\end{equation}

Note that $\d_1 \w^a_{\ b}$ can be extracted from
(\ref{varcon11}) using the matrix $M$ given in (\ref{defm}). The
result is:

\begin{equation}\label{nouvva}
\d_1 \w_{ab}=\d_2 \w_{2ab} +
\frac{1}{2}\left(T_{a,bc}-T_{c,ab}+T_{b,ca} \right)\eb^{c}
\end{equation}
where we defined $T_{c,ab}:=\frac{2}{D-2} \eta_{c[a}
\tilde{T}^{d}_{\ b]d}+\tilde{T}_{cab}$ with
$\tilde{T}^{\mu}_{\ ab}:=\delta_{2}\eb^d_{\rho}\Delta \omega_{\nu\
[a}^{\ c} \delta^{\mu \nu \rho}_{b]cd}$.

Of
course, the above result can also be obtained 
straightforwardly from equation (\ref{delome00}).

\subsection{The supersymmetry transformation law for the field
strength}\label{fielsusy} 

If the supergravity Lagrangian depends also on some (non)-Abelian
$p$-form $\Ab$, we may wish to give a supersymmetry transformation law
for the associated auxiliary field, namely the field strength $\Fb$.
Using equation (\ref{sugras}) in (\ref{varlag2}), this transformation
law follows from (see also the recipe given in the previous example):

\begin{equation}\label{fsvari}
-\frac{1}{2}\delta_{1}\left(\D\Fb \we \st \D\Fb \right)=0
\end{equation}
with $\D\Fb=\Fb -\Fb_{2}$, $\Fb_{2} (\fb)$ being the induced field
strength constructed from the dynamical fields $\fb$ (for instance
$\Fb_{2}=\ex \Ab + \dots $) through the equation of motion:
 $\left. \frac{\delta
\L_{\sug}}{\delta \Fb} \right|_{\Fb =\Fb _{2}}=0$.

Equation (\ref{fsvari}) gives the transformation law for $\Fb$:

\begin{equation}\label{trancoo}
\delta _{1} \Fb =\delta_{2} \Fb_{2} + \delta _{2}\eb^{a}\we \i_{a} \D
\Fb -\frac{\delta_{2} e}{2e} \D \Fb 
\end{equation}
where $\i_{a}$ denotes the interior product with respect to the vector
field $e^{\mu}_{a}$ (the inverse of the vierbein form
$\eb^{a}=e^{a}_{\mu} \ex x^{\mu}$) and $e$ the
determinant of the vierbein.

\bigskip 

To summarize: a $1^{st}$ order formulation of any supergravity with a
$1.5$ order action is given by the transformation laws (\ref{defd1}),
(\ref{nouvva}) and (\ref{trancoo}).

\subsection{Some useful trivial transformations}\label{utrss}

It may be preferable to  rewrite the transformation laws in a 
simpler form by twisting them slightly. The transformation
law for the gravitino can be generically written as:
\begin{equation}\label{trangrav}
\d_1 \bar{\pb} := \cex_{2} \bar{\e} + \mbox{\it more}
\end{equation}
where $\cex_{2}$ is the covariant derivative constructed
with the induced connection $\w^{ab}_{2}$ and {\it more} is the
remaining part of the 
supersymmetry transformation law, which does not contain any
derivative of
the transformation parameter $\bar{\e}$. 

Let us then define a new transformation law for the gravitino
(and for the connection), which differs from 
(\ref{trangrav}) by a term proportional to the equations of motion of
$\w^{a}_{\ b}$ (\ref{eqw1}) (this makes the Lorentz covariance more obvious
with $\cex_2 \rightarrow \cex$, $\cex$ being the covariant derivative
defined by the connection $\w^{a}_{\ b}$):

\begin{eqnarray}
\bar{\d}_1 \bar{\pb} &:=& \d_1 \bar{\pb}-\bar{\e} \D \w^{ab}
\frac{\g_{ab}}{4} =\cex \bar{\e} + \mbox{\it more}\label{trp1}\\ 
\bar{\d}_1 \w^{ab} &:=& \d_1 \w^{ab} + \Ob^{ab}\label{trw1}
\end{eqnarray}

\noindent and $\bar{\d}_1\mbox{\it (other fields)} := \d_1 \mbox{\it (other
fields)}$.

Here $\Ob^{ab}$ is the one form given as in equation
(\ref{twi2}). A direct way to compute it explicitly is the following:
$\Ob^{ab}$ has to be such that $\Ob^{ab}\we \frac{\delta
\L_{\sug}}{\delta \w^{ab}}-\bar{\epsilon} \D
\w^{ab}\frac{\gamma_{ab}}{4}\we \frac{\delta \L_{\sug}}{\delta
\bar{\pb}}=0$. Using the equation of motion of the connection
(\ref{eqw1}), we find that:

\begin{equation}\label{twist}
\Ob^c_{\ [a} \we \S_{b]c} = -\frac{\k^2}{2}\bar{\e}
\g_{ab} \frac{\d \L_{\sug}}{\d \bar{\pb}}
\end{equation}
This kind of twist may sometimes
 significantly simplify the final expression for
the connection transformation law.

\bigskip 

Finally, if the transformation law of the gravitino (\ref{trangrav})
depends on the induced field strength $\Fb_{2}$, it is again
possible to twist it in a similar way by adding some trivial gauge
transformation proportional to the equation of motion of the field
strength. The net result would be the change $\Fb_{2} \rightarrow \Fb$
in (\ref{trangrav}).

\section{The $1^{st}$ order ${\cal{N}}_4 =1$ supergravity}

As a first example let us now derive (systematically) the old result 
of Deser and Zumino \cite{DZ} for the ${\cal{N}}_4=1$ supergravity in 4 dimensions.
Let us first use the general method of the previous
section. That is, starting from a $1.5$ formulation we will derive the
connection supersymmetry transformation law and improve it by adding a
trivial gauge transformation. In the last subsection we shall present
a new method 
to obtain the same result in a more straightforward way. We shall use
this method in the next sections to treat both the  ${\cal{N}}_4=2$ and
the ${\cal{N}}_{11}=1$ supergravities.

\subsection{The Lagrangian and the equations of motion}\label{lageq}

Let us start with  the $1^{st}$ order Lagrangian of ${\cal{N}}_4 =1$ supergravity\footnote{The conventions are the
following (see also appendix A): 
$\eta^{ab}=\{ -,+,+,+\}$, $\e_{0123}=1$, $\g^a$ are four real Majorana 
matrices, $\g^5=\g^0 \g^1 \g^2 \g^3$, $\left(\g^5\right)^2=-\1$,  $4\k^2=16\pi G$, $G$ being the Newton 
constant.} (as given by the so-called $1.5$ formalism):

\begin{equation}\label{lag1}
\L_1= -\frac{1}{4\k^2} \R^{ab} \we \S_{ab} + \frac{i}{2} \bar{\pb}\we \g^5 \gab_{(1)} \we \cex \pb
\end{equation}

Where $\gab_{(1)} := \g_a \eb^a$, $\cex \pb := \ex \pb + \frac{\g_{ab}}{4} \w^{ab} \we \pb$ with $\g_{a_1
\ldots a_r} := \g_{[a_1} \ldots \g_{a_r]}$ and the definitions of
$\R^{ab}$  and $\S_{ab}$ were given in the previous section. The equations of 
motion corresponding to Lagrangian (\ref{lag1}) are:

\begin{eqnarray}
\frac{\d \L_1}{\d \eb^a} &=& -\frac{1}{4 \k^2} \R^{bc} \we \S_{bca}
-\frac{i}{2} \bar{\pb} \g^5 \g_a \we \cex \pb = 0 \label{eqmoe1}\\
\frac{\d \L_1}{\d \bar{\pb}} &=& i \g^5 \gab_{(1)} \we \cex
\pb-\frac{i}{2} \g^5 \g_a \left( \X^a - \frac{i \k^2}{2} \bar{\pb}
\g^a \we \pb \right) \we \pb = 0 \label{eqmop1}\\
\frac{\d \L_1}{\d \w^{ab}} &=& -\frac{1}{4 \k^2} \left( \X^c - \frac{i
\k^2}{2} \bar{\pb} \g^c\we \pb \right) \we \S_{abc} = 0 \label{eqmow1}
\end{eqnarray}
with $\X^a:=\ex \eb^{a} +\w^{a}_{\ b}\we \eb ^{b}$. 

Note that the very last term of equation (\ref{eqmop1}), namely
$-\frac{\kappa ^{2}}{4}\gamma^{5}\gamma _{a}\pb \we (\bar{\pb}
\g^a \we \pb)$, vanishes
identically after a Fierz rearrangement. It has been however added to
(\ref{eqmop1}) to emphasize the vanishing of the sum of the
last two terms after
making use of the equations of motion of the connection (\ref{eqmow1}).

\subsection{The supersymmetry transformation law}\label{thsussy}

The Lagrangian (\ref{lag1}) is invariant under local supersymmetry as a $2^{nd}$
order functional of $\eb^a$ and $\pb$. That is, if we use the equations of motion of the 
connection $\w^{ab}$, the following transformation laws leave the action 
invariant:
\begin{eqnarray}
\d_2 \bar{\pb} &=& \cex_2 \bar{\e} = \ex \bar{\e} - \bar{\e} \w_2^{ab}
(\eb,\pb) 
\frac{\g_{ab}}{4} \label{transp2}\\
\d_2 \eb^a &=& i \k^2 \bar{\e} \g^a \pb \label{transe2}
\end{eqnarray}

\bigskip 

A purely $1^{st}$ order symmetry of supergravity is defined by $\d_1 \eb^a :=
\d_2 \eb^a$ and $\d_1 \bar{\pb} :=  \d_2 \bar{\pb}$, plus the
supersymmetry transformation law for the connection given by 
(\ref{varcon11}). 
%To proceed, we need the supersymmetry transformation
%induced on the $2^{nd}$ order connection, namely $\d_{0} \w_2^{ab}
%(\eb^a, \pb)$. 
Using the fact that $\w_2^{ab}$ identically satisfies
equation (\ref{eqmow1}), we can compute $\d_{1} \w_2^{ab}(\eb^a, \pb)$
in a simple way.
%\begin{equation}\label{induced}
%\cex_{2} \d_{0} \eb^{a} + \d_{0} \w^{a}_{2 b}\we \eb^{b} +i\k^{2}
%\d_{0}\bar{\pb} \g^{a} \we \pb =0
%\end{equation}
%Again, $\cex_{2}$ is the covariant derivative associated to the
%$2^{nd}$ order connection $\w^{a}_{2 b}$. Now using formulas
%(\ref{transp2}-\ref{transe2}), equation (\ref{induced}) simplifies
%to:
%\begin{equation}\label{induce2}
%\d_{0} \w^{a}_{2 b}\we \eb^{b}=i\k^{2} \bar{\e} \g^{a} \cex_{2} \pb
%\end{equation}
%\begin{equation}\label{interm}
%\d_{0} \w^{c}_{2 [a}\we \S_{b]c}=\frac{i\k^{2}}{2}
%\bar{e}\g^{c}\cex_{2}\pb \we \S_{abc}
%\end{equation}
%Now, wedging the above equation with $\S_{abc}$ and using the identity
%$\eb^b\we\S_{a_1 \ldots a_r}=(-)^{r+1}r
%\d^b_{[a_1} \S_{a_2 \ldots a_r]}$, 
Equation (\ref{varcon11}) becomes
after some algebra:

\begin{equation}\label{varw11}
\d_1 \w^c_{\ [a} \we \S_{b]c} =  -\frac{i\k^2}{2} \bar{\e} \g^c \cex \pb
\we \S_{abc}-\Ab_{ab} + \Bb_{ab} +\Cb_{ab} 
\end{equation}
where we defined,
\begin{eqnarray}
\Ab_{ab} &:=& \frac{1}{4} \D \w^c_{\ d}\we\eb^d\we\d_2\S_{abc}\label{defa}\\
\Bb_{ab} &:=& \frac{i\k^2}{2}  \bar{\e}\g^5 \g_{[a} \D
\w_{b]c}\we\eb^c\we\pb\label{defb}\\
\Cb_{ab}&:=& \frac{i\k^{2}}{4}\bar{\e}\g^{5}\gab \we \D \w_{ab} \label{defc}
\end{eqnarray}

Note that we just found an explicit expression
 for $\d_1 \w^c_{\ [a}\we \S_{b]c}$, which can (always) be inverted
 (as equation (\ref{eqw1})) in any
 spacetime dimension $D \geq 3$ using the matrix (\ref{defm}).
Finally,
 we see from (\ref{varw11}) that there is no term proportional to the
 derivative of the gauge parameter (namely $\ex \bar{\e}$) in the
 transformation law of the connection (in other words the connection
 remains supercovariant). This is due to two facts:
\begin{enumerate}
\item [] $\bullet$ The {\it induced} connection 
$\w^{ab}_{2}(\eb,\pb)$ was in this case effortlessly supercovariant, 
its supercovariance is still obscure though and
 it may seem that it is a four-dimensional  accident  \cite{VN};
 more precisely it is only in four dimensions that one may require
simultaneously the independence of the connection and the absence of
fermionic quartic terms in the action, in fact here we shall always use the
supercovariant connection as the independent variable by using the
freedom mentioned at the end of section 2.

\item [] $\bullet$ The last term of the rhs of (\ref{varcon11}) (or of
(\ref{nouvva})) is
also supercovariant since it depends on $\bar{\epsilon}$ only through
$\delta_{2} \eb^{a}$ (which is independent of $\partial_{\mu}
\bar{\epsilon}$, see (\ref{transe2})). 
\end{enumerate}

We may pause at this stage to reflect on the general structure of
supersymmetry transformation laws. The anticommutator of two of them
gives a diffeomorphism consequently one has the typical multiplet
structure $(B,F,A)$
with B bosonic, F fermionic and A auxiliary ie nonpropagating fields.
Their
variations are schematically
\begin{eqnarray}
\delta B &\propto& F \\
\delta F &\propto& A + \partial B \\
\delta A &\propto& \partial F 
\end{eqnarray}

If one insists on a geometrical approach one is led to replace the
term
$ \partial B$ by the Lorentz connection in the case of the graviton
multiplet
but this uses the equation of motion for the  connection so first
order formalisms and offshell supersymmetry algebra are somewhat antinomic. 
(W. Siegel informed us recently that he
has made progress on this issue).

Nevertheless supercovariance is useful to compute the superpotential
corresponding to the 
 supersymmetry gauge invariance in $1^{st}$ order formalism, see
\cite{HJS}. 

At this point the definition of $\d_1 \w^{ab}$ given by
(\ref{varw11}), together with
(\ref{transp2}-\ref{transe2}) provide a $1^{st}$ order formulation of 
supergravity. It is however preferable to add two trivial gauge
transformations to simplify the previous expressions:

\begin{enumerate}
\item The last term of equation (\ref{varw11}), namely $\Cb_{ab}$
(\ref{defc}), can be eliminated. In
fact it corresponds to a trivial gauge transformation of the type
(\ref{entriv}). This can be checked as follows: due to equation
(\ref{eqw1}), we have that $\delta_{1} \w^{ab}\we \frac{\delta
\L_{1}}{\delta \w^{ab}}=-\frac{1}{2\kappa ^{2}}\D \w^{ab}\we\delta_{1}
 \w^{c}_{\ a}\we \S_{bc}$. Then, $\Cb_{ab}$
contributes with a term proportional to $\D \w^{ab} \we \D \w_{ab}$,
which vanishes identically (the connection is a one-form).
\item The second trivial gauge transformation was  
explained
in subsection \ref{utrss} through equations
(\ref{trangrav}-\ref{twist}). It allows to rewrite the gravitino
transformation law in terms of the $1^{st}$ order covariant
derivative $\cex$ (instead of $\cex_{2}$):

\begin{eqnarray}
\bar{\delta}_{1} \bar{\pb} &:=& \cex\bar{\epsilon}\label{bdel1}\\
\bar{\delta}_{1}\w^{ab} &:=& \delta_{1}\w^{ab}+\Ob^{ab}\label{bdel2}
\end{eqnarray}

The one-form $\Ob^{ab}$ is computed from (\ref{twist}) using the gravitino equation of
motion (\ref{eqmop1}). After some
rearrangement, we find  that:

\begin{equation}\label{omb}
\Ob^c_{\ [a} \we \S_{b]c} = -i\k^2 \bar{\e} \g^5 \g_{[a} \eb_{b]}
\we \cex \pb +\frac{i\k^2}{2} \bar{\e} \g^c \cex \pb \we
\S_{abc}+\Ab_{ab}-\Bb_{ab}
\end{equation}
Where $\Ab_{ab}$ and $\Bb_{ab}$ were respectively defined in
(\ref{defa}) and (\ref{defb}).

The twisted $1^{st}$ order supersymmetric variation of the connection
is then given simply by combining (\ref{trw1}) together with (\ref{varw11})
and (\ref{omb}):

\begin{equation}\label{nouvw}
\bar{\d}_1 \w^c_{\ [a} \we \S_{b]c} = -i\k^2 \bar{\e} \g^5 \g_{[a} \eb_{b]} \we \cex \pb
\end{equation}

%It is well known (see \c{H3}) that $\d_1$ will still be a gauge symmetry of the Lagrangian $\L_1$ (\ref{lag1}) if we modify the variation of the connection by:

%\B \d_1 \omega^{\n cd} := \d_2 \omega^{\n cd} + \bar{\e} M_{\m ab}^{\n cd} 
%\frac{\g^{ab}}{4} \frac{\d L_1}{\d \psi_\m} \lb{trwf} \E
Note that all the terms proportional to the equations of motion of the
connection (through $\D \w^{ab}$), namely $\Ab_{ab}$ and $\Bb _{ab}$ disappear.

One finally recovers the old result of Deser and Zumino \cite{DZ} after inverting
(\ref{nouvw}) using the matrix (\ref{defm}) (as for (\ref{eqw1})):

\begin{equation}\label{varomf}
\bar{\d}_1 \w_{ab} = 2 i\k^2 \bar{\e} \g^5 \left(\gab_{(1)} \st\psi_{ab} 
-\frac{1}{2} \eb_a \g^c \st\psi_{cb}
+ \frac{1}{2} \eb_{b} \g^c \st\psi_{ca} \right) ,
\end{equation}
where $ \st\psi_{\m\n} =\frac{1}{2} \e_{\m\n\r\sigma} D^\r
\psi^{\sigma}$.

\end{enumerate}

Thus, the first order formulation of ${\cal{N}}_4 =1$
supergravity is given by the Lagrangian (\ref{lag1}) and the locally
supersymmetric transformation laws  (\ref{transe2}), (\ref{bdel1}) and
(\ref{varomf}) (or in a less aesthetical way by (\ref{transp2}),
(\ref{transe2}) and (\ref{varw11})). But the relative simplicity of
(\ref{varomf}) (comparable to that of $\delta_{2}\w ^{ab}_{2} (\eb,
\pb)$) remains mysterious.

\subsection{A new method}\label{newm}

Our idea now is to formulate better the choice of the twist (trivial gauge
transformation) and then to rewrite the equation for
$\bar{\delta}_{1} \w^{ab}$. Geometrically, one would like to keep
(\ref{transe2}) and (\ref{bdel1}) and find the $\bar{\delta}_{1}
\w^{ab}$ of  (\ref{varomf}) in a more
straightforward and less surprising way. 

Let us begin with a general
discussion: 
suppose that our Lagrangian is invariant under a gauge
symmetry given by:

\begin{equation}\label{tranlf}
\delta_{\xi} \fib^{i} = \ex \xi^{a} \we \Db^{i}_{a} +
\xi^{a}\tilde{\Db}^{i}_{a}  
\end{equation}
where $\fib^{i}$ is now a $p_{i}$-form which goes for all the fields (even
auxiliary) of the theory. 

Moreover, we suppose that the theory is a first order one, in the sense
that $\delta_{\xi} \fib^{i}$ and $\Eb_{i}:=\frac{\delta \L}{\delta
\fib^{i}}$ depend on $\fib^{i}$ and $\ex \fib^{i}$ (and no more
derivatives of the fields). Let us now define
the $(D-1)$-form:

\begin{equation}\label{defw}
\Wb_{\xi}:=\xi ^{a} \Db^{i}_{a} \we \Eb_{i}
\end{equation}

Then, the following identity holds (offshell of course):

\begin{equation}\label{idensi}
\frac{\delta \Wb_{\xi}}{\delta \fib^{i}} = \frac{\partial}{\partial
\ex \fib^{i}}\left( \delta_{\xi} \fib^{j}\we \Eb_{j}\right)
\end{equation}
where $\frac{\delta }{\delta \fib^{i}}=\frac{\partial }{\partial
\fib^{i}}- (-)^{p_{i}} \ex \frac{\partial }{\partial \ex \fib^{i}}$.

The proof is given in \cite{Si} (see also \cite{SiT}). 
The point is now that the identity (\ref{idensi}) can be used to
compute straightforwardly $\bar{\delta}_{1} \w^{ab}$ from
$\bar{\delta}_{1} \eb^{a}$ and $\bar{\delta}_{1} \pb$. 
Before starting concrete calculations, we will comment on the general
meaning of equation (\ref{idensi}).

There is a fundamental reason  why (\ref{idensi}) is a useful
formula for computing transformation laws. In fact, it is the
Lagrangian (covariant) version of a standard formula of
Hamiltonian formalism. Namely in the Hamiltonian formalism, a
transformation law $\delta_{\xi} \Fib^{I}$ can
be computed from the 
charge $Q_{\xi}$ and the symplectic structure $\Omega_{IJ}$ by the following formula:

\begin{equation}\label{hacha2}
\delta_{\xi} \Fib^{I} = \left\{Q_{\xi},\Fib^{I}
\right\}_{PB}
\end{equation}
where now $\Fib^{I}$ are phase space fields. The Poisson bracket (PB)
is defined by the inverse of the symplectic structure by
$\left\{, \right\}:=\frac{\delta }{\delta
\Fib^{I}}\Omega^{IJ}\frac{\delta }{\delta \Fib^{J}}$.

Then, equation (\ref{hacha2}) can be rewritten as:
\begin{equation}\label{hacha}
\frac{\delta Q_{\xi}}{\delta \Fib_{I}} = \delta_{\xi} \Fib^{J}\Omega_{JI} 
\end{equation}

Although (\ref{idensi}) and (\ref{hacha}) look different, they are in
some way analogous. First of all the lhs of (\ref{idensi}) is nothing
but the Euler-Lagrange variation of the Noether current $\J_{\xi}$ (and
so the local covariant version of $Q_{\xi}$ of (\ref{hacha})). This is a direct
consequence of the fact that $\J_{\xi}=\Wb_{\xi}+\ex \U$ ($\U$
being the superpotential) \cite{JS}. Now, the rhs of (\ref{idensi})
contains two terms. The first one, $\delta_{\xi}
\fib^{j}\we\frac{\partial}{\partial 
\ex \fib^{i}}   \Eb_{j}$, corresponds to $\delta_{\xi}
\Fib^{J}\Omega_{JI}$ of (\ref{hacha}). In fact, $\frac{\partial}{\partial 
\ex \fib^{i}}   \Eb_{j}$  is antisymmetric in $i$ and $j$ and is a
local version (non integrated) of the Witten-Crnkovic-Zuckerman 2-form
\cite{WCZ}(see \cite{Si}, \cite{SiT}) times $\delta_{\xi}
\fib^{j}$. The second term, namely $\frac{\partial}{\partial 
\ex \fib^{i}} \delta_{\xi}
\fib^{j} \we \Eb_{j}$, vanishes on-shell. It is not so disturbing
to find such an additional contribution in our covariant formulas
since it also exists in the Hamiltonian formalism; in fact, the
equivalence between (\ref{hacha}) and the 
symmetry of the original action is only up to some identifications
of the type ``$\dot{q}=p$''. In our case, this kind of identifications
are part of the equations of motion ($1^{st}$ order theories).

\bigskip

Let us come back to our ${\cal{N}}_4=1$ supergravity 
and use the identity (\ref{idensi}) to compute the
supersymmetry transformation law of the connection. First of all, we
will comment  
on the following points derived in previous sections:

\begin{enumerate}
\item [] $\bullet$ By using some trivial gauge transformation, it is always possible to rewrite the supersymmetry
transformation law of the gravitino in such a way that it depends on the
independent field $\w^{ab}$ instead of the induced auxiliary field $\w^{ab}_{2}
(\fb)$. The mathematical consequence of this twist is that now
$\frac{\partial \bar{\delta}_{1}\bar{\pb}}{\partial \partial_{\mu}
e^{a}_{\nu}}=0$ (see the definition (\ref{bdel1})). That was obviously
false for $\delta_{2} 
\bar{\pb}$ of (\ref{transp2}).

\item [] $\bullet$ The induced connexion $\w^{ab}_{2}
(\fb)$ has been chosen to be supercovariant in the sense
that its supersymmetry variation does not depend on the derivative of
the gauge parameter $\bar{\epsilon}$. In that case, the independent
connexion  $\w^{ab}$ will also be supercovariant. This is a
consequence of the general result (\ref{varcon11}-\ref{nouvva}) (itself following
rather strong hypothesis on the gravitational kinetic term) and of the fact
that both the supersymmetry transformation law of the vierbein
(\ref{transe2}) and the
trivial gauge transformations (\ref{trw1}-\ref{twist}) do not depend on the derivative of the
gauge parameter $\partial_{\mu}\bar{\epsilon}$. 
\end{enumerate}
Two conclusions followed:
\begin{enumerate}
\item there should exist some $\bar{\delta}_{1}\w^{ab}$, which,
together with $\bar{\delta}_{1}\bar{\pb}=\cex\bar{\epsilon}$ and
$\bar{\delta}_{1}\eb^{a}=i\kappa^{2}\bar{\epsilon}\gamma^{a}\pb$,
leave the supergravity action (\ref{lag1}) invariant.
\item this  $\bar{\delta}_{1}\w^{ab}$ does not depend on the
derivatives of the parameter $\bar{\epsilon}$.
\end{enumerate}

We are now ready to use the identity (\ref{idensi}). From the previous
discussion, the gravitino is the only field whose
supersymmetry transformation law depends on the derivative of
$\bar{\epsilon}$. Thus, the definitions  (\ref{tranlf}) and (\ref{defw}) give:

\begin{equation}\label{wsusy}
\Wb_{\bar{\epsilon}}=\bar{\epsilon}\frac{\delta \L_{1}}{\delta
\bar{\pb}}
\end{equation}

Replacing the equation (\ref{eqmop1}) in the above expression,
we compute:

\begin{equation}\label{wvariee}
\frac{\delta \Wb_{\bar{\epsilon}}}{\delta
\eb^{a}}=\frac{i}{2}\bar{\epsilon}\gamma^{5} \gamma_{a} \cex\pb
-\frac{i}{2} \cex\bar{\epsilon}\we \gamma^{5} \gamma_{a} \pb
\end{equation}

According to the identity (\ref{idensi}), this expression is equal to

\begin{equation}\label{equto}
\frac{\partial}{\partial \ex \eb^{a}}\left(\bar{\delta}_{1}\eb^{b}\we
\frac{\delta \L_{1}}{\delta \eb^{b}}+\bar{\delta}_{1}\bar{\pb }\we
\frac{\delta \L_{1}}{\delta \bar{\pb }}+\bar{\delta}_{1} \w^{cd}\we
\frac{\delta \L_{1}}{\delta \w^{cd} } \right).
\end{equation}

Using the known expressions for $\bar{\delta}_{1}\eb^{b}$ (\ref{transe2}),
$\bar{\delta}_{1}\bar{\pb }$ (\ref{bdel1}) (which do not depend on the derivatives of
the vierbein, see the first comment above) and the equations of motion
(\ref{eqmoe1}-\ref{eqmop1}), our expression becomes:

\begin{equation}\label{simplif}
-\frac{i}{2}\cex\bar{\epsilon}\we \gamma ^{5} \gamma_{a}\pb +
\frac{\partial}{\partial \ex \eb ^{a}} \left( \bar{\delta}_{1} \w^{cd}\we
\frac{\delta \L_{1}}{\delta \w^{cd}}  \right)
\end{equation}

\noindent So equations (\ref{wvariee}) and (\ref{simplif}) (together with
(\ref{idensi})) imply:

\begin{equation}\label{npuid}
\frac{\partial}{\partial \ex \eb ^{a}} \left( \bar{\delta}_{1} \w^{cd}\we
\frac{\delta \L_{1}}{\delta \w^{cd}}  \right) = \frac{i}{2}\bar{\epsilon}\gamma^{5} \gamma_{a} \cex\pb
\end{equation}

\bigskip

This kind of equation will reappear in a more complicated form in the
case of ${\cal{N}}_4=2$ or ${\cal{N}}_{11}=1$ supergravities, that
is why we shall now investigate it in general. Let us denote by
$\omega ^{i}$ the auxiliary fields, $E_{i}$ their associated equations of
motion and by $x^{\alpha}$ the $\ex \eb ^{a}$'s (thus $\alpha$ stands
for two spacetime indices and one Lorentz index). Then, equation
(\ref{npuid}) can be rewritten as:

\begin{equation}\label{rrro}
\partial_{\alpha}\left( \bar{\delta}_{1}\omega^{i} E_{i}\right)= R_{\alpha}
\end{equation}

In our case, the equations $E_{i}$ are linear in $x^{\alpha}$ (that
is, the equations of the connection are linear in the derivatives of the
vierbeins). The $x^{\alpha}$ dependence can be expressed as
$E_{i}=E_{i\alpha }x^{\alpha }+E_{i0}$. On the other hand from
(\ref{npuid}) one reads that $R_{\alpha}$ is independent of $x^{\alpha}$. In
that case, we can {\it choose} $\bar{\delta}_{1}\omega^{i}$ to be also
independent of $x^{\alpha}$, and then (\ref{rrro}) implies that:

\begin{equation}\label{ddd}
\bar{\delta}_{1}\omega^{i}  E_{i\alpha } = R_{\alpha}
\end{equation}

We used the word {\it choose} because (\ref{rrro}) determines
$\delta_{1}\omega^{i}$ only up to some trivial gauge symmetry of the
type (\ref{entriv}), which can eventually depend on $x^{\alpha}$. The
important point is that $\bar{\delta}_{1}\omega^{i}$ independent of
$x^{\alpha}$ is consistent with (\ref{npuid}). The last step is to
invert equation (\ref{ddd}) to extract
$\bar{\delta}_{1}\omega^{i}$. In our case, this is possible because
the indices $i$ and $\alpha$ have the {\it same} dimension, namely
$\frac{D^{2} (D-1)}{2}$. This is due to the fact that $\w^{ab}$
and
$\eb ^{a}$  are, in a covariant sense, canonically conjugate
fields. 

Thus, using the first identity of
 (\ref{eqmow1}), we can find a solution for $\bar{\delta}_{1}
\w^{cd}$ (from (\ref{ddd})): 

\begin{equation}\label{solw}
-\frac{1}{4\kappa ^{2}}\bar{\delta}_{1} \w^{cd}\we \S_{acd} =
\frac{i}{2}\bar{\epsilon}\gamma^{5}\gamma_{a}\cex\pb
\end{equation}

This last equation is equivalent to (\ref{nouvw}). This can be checked
by ``wedging'' the above expression with $\eb _{b}$ and
antisymmetrising in the $a$ and $b$ indices. Then, it can be inverted
to extract $\bar{\delta}_{1} \w^{ab}$. The result is of course again
given by (\ref{varomf}). 

This new method will be used in the following sections to treat the
${\cal{N}}_4=2$ and ${\cal{N}}_{11}=1$ supergravities.

\section{$1^{st}$ order formalism for  ${\cal{N}}_4=2$ supergravity}

Contrary to the ${\cal{N}}_4=1$ case, there seems to be
 no published first order
formalism for ${\cal{N}}_4=2$ supergravity. We will give it here.

\subsection{The $1^{st}$ order Lagrangian}

The Lagrangian of ${\cal{N}}_4=2$ supergravity in D=4 dimensions
depends on the  
one-form vierbein $\eb^a$, the associated one-form so(1,3) (spin) connection 
$\w^a_{\ b}$, two Rarita-Schwinger Majorana spinor-one-forms the gravitinos 
$\pb^A$ ($A=1,2$ is the internal global $SO(2)$ R-symmetry index) and one 
Abelian one-form gauge connection, the photon $\Ab$, and its
associated field strength $\Fb$. 
The 4-form Lagrangian is given by\footnote{The conventions are the
same as for the previous subsection (see also appendix A) together with
the antisymmetric 
SO(2)-invariant tensor  $\ve^1_{\ 2}=-\ve^{2}_{\ 1}=1$ and $\ve_{\
C}^A \ve_{\ B}^C=-\delta^{A}_{\ B}$.} \c{VN}:

\begin{eqnarray}
\L_2 &=& -\frac{1}{4\k^2} \R^{ab} \we \S_{ab} + \frac{i}{2} \bar{\pb}_A\we \g^5 \gab_{(1)} \we \cex \pb^A - \frac{1}{2} \Fb \we \st \Fb \nonumber\\
&\ & +\left(\st \Fb -\bb  \right)\we \left(\ex \Ab -\ab
\right)-\frac{1}{2}\ab \we \bb\label{lagsugra} 
\end{eqnarray}
with the definitions 

\begin{equation}\label{defab}
\ab:= \frac{i \k}{2}\ve_{\ B}^A \bar{\pb}_A \we\pb^B \hskip.9cm
\bb:=  \frac{i \k}{2}\ve_{\ B}^A \bar{\pb}_A \gamma ^{5}\we\pb^B
\end{equation}

%\B +\frac{i \k}{4} \ve_{\ B}^A \bar{\pb}_A \we \left(\st \Fb +\st \hat{\Fb} - \g^5 \left(\Fb +\hat{\Fb} \right) \right) \we \pb^B \lb{lagsugra} \E

Notations, conventions and useful formulas can be found in appendix A.
Moreover, the $\st$ symbol refers to standard Hodge-duality.

\bigskip

The equations of motion of (\ref{lagsugra}) are:

\begin{eqnarray}
\frac{\d \L_2}{\d \eb^a} &=& -\frac{1}{4 \k^2} \R^{bc} \we \S_{bca}
-\frac{i}{2} \bar{\pb}_A \g^5 \g_a \we \cex \pb^A + \t_{a} = 0
 \label{eqmoe}\\
\frac{\d \L_2}{\d \bar{\pb}_A} &=& i \g^5 \gab_{(1)} \we \hat{\cex}
\pb^A-\frac{i}{2} \g^5 \g_a \pb^A \we\left( \X^a - \frac{i \k^2}{2} \bar{\pb}_B
\g^a \we \pb^B \right)  \nonumber\\
&\ & + i\kappa \ve^A_{\ B}\gamma ^{5} \pb^{B}\we \left(\Fb - \ex \Ab
+\ab  \right) = 0 \label{eqmop}\\
\frac{\d \L_2}{\d \Ab} &=& \ex \left(\st \Fb-\bb\right) = 0 \label{eqmoa}\\
\frac{\d \L_2}{\d \w^{ab}} &=& -\frac{1}{4 \k^2} \left( \X^c - \frac{i
\k^2}{2} \bar{\pb}_A \g^c\we \pb^A \right) \we \S_{abc} = 0 \label{eqmow}\\
\frac{\d \L_2}{\d \Fb} &=& \st \left(\ex \Ab -\ab -\Fb \right) = 0 \label{eqmof}
\end{eqnarray}

Let us comment the above equations:

\begin{enumerate}
\item [] $\bullet$ The last term in the rhs of  (\ref{eqmoe}), namely
$\t_{a}$, denotes the electromagnetic energy-momentum tensor and is
given\footnote{A more explicit formula can be given for $\t_{a}$ by
using that
$\frac{\partial}{\partial \eb^{a}} (\Fb \we \st \Gb)= (-)^{p}\Fb \we
\i_{a} (\st \Gb)-\i_{a}\Fb \we \st \Gb $, for $\Fb$ and $\Gb$ any
$p$-forms independent of $\eb^{a}$, see \cite{Thi}, \cite{SiT}.} by $\t_{a} = \frac{\partial}{\partial\eb_{a}} (\st \Fb\we (\ex \Ab -\ab
-\frac{1}{2}\Fb ) )$. 
\item [] $\bullet$ The supercovariant derivative appearing in the
gravitini's equations of motion (\ref{eqmop}) is defined by:
$\hat{\cex} \pb^A:=
\cex \pb^A - \frac{\k}{2} \ve^A_{\ B}(\Fb_a \g^a +\st \Fb_a \g^a \g^5)
\we \pb^B$, with $\Fb_a:= \i_{\eb_a} \Fb=F_{ab} \eb^b$ and $\st
\Fb_a:= \i_{\eb_a}(\st \Fb)=(\st F)_{ab} \eb^b$. To compute the result
(\ref{eqmop}), the  following two identities are useful:
\begin{enumerate}
\item [] the Fierz rearrangement 
\[
\g_a \pb^A\we( \bar{\pb}_B\g^a\we\pb^B)+\ve^A_{\ B} \pb^B\we
( \ve^C_{\ D} \bar{\pb}_C\we \pb^D) + \ve^A_{\ B}\g^5\pb^B\we
( \ve^C_{\ D} \bar{\pb}_C \g^5\we \pb^D)=0
\]

\item [] a supercovariant derivative identity 
\[
i\gamma ^{5}\gab_{(1)}\we \hat{\cex} \pb ^{A}=i\gamma
^{5}\gab_{(1)}\we \cex \pb ^{A} -i\kappa\left(\st \Fb +\gamma
^{5}\Fb \right)\we \ve^A_{\ B} \pb ^{B}.
\]

\end{enumerate}

Note also that the last two terms of (\ref{eqmop}) vanish by making
use of the auxiliary fields equations of motion (\ref{eqmow}-\ref{eqmof}).

\item [] $\bullet$ The unique solution to the field
strength equation of motion, namely
$\Fb_{2}=\ex \Ab -\ab$ is the so-called supercovariant field strength, in the
sense that its induced supersymmetry transformation law (see equations
(\ref{tranp}-\ref{trana}) below) does not depend on the derivatives of
the gauge parameter\footnote{The same happens for the induced spin
connexion as in the ${\cal{N}}_4=1$ case, namely it is already supercovariant.} $\bar{\epsilon}$. If we
eliminate the auxiliary field $\Fb$, we
recover the so-called ``$1^{st}$'' order formulation, where the
gravitational part of the Lagrangian is really  $1^{st}$ order whereas
the Maxwell part is not.
\item [] $\bullet$ The Bianchi identity and equation of motion for the
$U(1)$ gauge field read symmetrically:
$d(\Fb_{2}+ \ab) = 0$ and $d(\st \Fb_{2} -\bb) = 0$. 
The supercovariantisation term in  the field strength is dual to the
fermionic ``source" in its equation of motion. Both quadratic
expressions in the fermions should be rewritten (using the
transformation rules given in the next section) as part of 
 supercovariant differentiation.
\end{enumerate}

\subsection{The supersymmetry transformation laws}\label{susytr}

The $2^{nd}$ order Lagrangian which is obtained from (\ref{lagsugra})
after eliminating the auxiliary fields $\w^{ab}$ and $\Fb$ is
invariant under the following supersymmetry transformation laws:

\begin{eqnarray}
\d_2 \bar{\pb}_A &=&  \hat{\cex}_{2} \bar{\e}_A := \cex_{2} \bar{\e}_A
-\frac{\k}{2} \ve^{B}_{\ A} \bar{\e}_B (\hat{\Fb}_{2a} \g^a - \st
\hat{\Fb}_{2a}  \g^5\g^a)\label{tranp}\\
\d_2 \eb^a &=& i \k^2 \bar{\e}_A \g^a \pb^A\label{trane}\\
\d_2 \Ab &=& i \k \ve_{\ B}^A \bar{\e}_A \pb^B\label{trana}
\end{eqnarray}

As before, the index $_{2}$ in the rhs of (\ref{tranp}) indicates that the auxiliary fields are
the {\it induced} ones, namely $\cex_{2}=\cex (\w^{ab}_{2}(\eb , \pb 
))$ and $\Fb_{2}=\ex\Ab -\ab$.

The purpose is now to extend the $2^{nd}$ order supersymmetry
transformation laws (\ref{tranp}-\ref{trana}) to the $1^{st}$ order
Lagrangian (\ref{lagsugra}). To proceed, we will use the new method
presented in section \ref{newm}. Let us first emphasize three points:

\begin{enumerate}
\item We can already give a $1^{st}$ order formulation for the
${\cal{N}}_4=2$ supergravity by $\delta _{1}\bar{\pb}_{A}=\delta
_{2}\bar{\pb}_{A}$, $\delta_{1}\eb^{a}=\delta_{2}\eb^{a}$,
$\delta_{1}\Ab=\delta _{2}\Ab$ (equation (\ref{defd1})) and by
equations (\ref{nouvva}) and (\ref{trancoo}) for the transformation
law of the auxiliary fields.
\item For our choice of action the 
{\it induced} auxiliary fields $\w^{ab}_{2}$ and $\Fb_{2}$
are supercovariants, in the sense that their supersymmetry
transformation laws (induced by (\ref{tranp}-\ref{trana})) do not
contain derivatives of the gauge parameter (i.e. no terms
proportional to $\partial_{\mu}\epsilon$).
\item There should exist some transformation laws
$\bar{\delta}_{1}\w^{ab}$ and $\bar{\delta}_{1}\Fb$ which,
together with 
\begin{eqnarray}
\bar{\delta}_{1} \bar{\pb}_{A} &:=& \hat{\cex}
\bar{\epsilon}_{A}\label{ndet1}\\
\bar{\delta}_{1} \eb^{a} &:=& \delta_{2}\eb^{a}\label{ndet2}\\
\bar{\delta}_{1} \Ab &:=& \delta_{2}\Ab\label{ndet3}
\end{eqnarray}
leave the Lagrangian (\ref{lagsugra}) invariant (up to some surface
term). In fact, $\bar{\delta}_{1}$ differ from the $\delta_{1}$ (see
the above point 1.) by some trivial gauge transformations (which allow
$\w^{ab}_{2}\rightarrow \w^{ab}$ and $\Fb_{2}\rightarrow \Fb$ in
(\ref{tranp})) of the type (\ref{twi1}-\ref{twi2}).  
\end{enumerate}

To compute simultaneously $\bar{\delta}_{1}\w^{ab}$ and
$\bar{\delta}_{1}\Fb$, we will use the identity
(\ref{idensi}). First of all, these two transformation laws will
differ from $\delta_{2}\w^{ab}_{2}$ and
$\delta_{2}\Fb_{2}$ by some terms which are proportional to
$\bar{\epsilon}$ (and which vanish on-shell). Thus the gravitini are the only fields
whose supersymmetry transformation law contains terms proportional to
$\partial_{\mu}\bar{\epsilon}$. In that case, the $(D-1)$-form defined
by (\ref{defw}) becomes:

\begin{equation}\label{wpour2}
\Wb_{\epsilon}=\bar{\epsilon}_{A} \frac{\delta \L_{2}}{\delta \bar{\pb}_{A}}
\end{equation}

Using then the first identity in (\ref{eqmop}), we can compute the
 following $(D-2)$-forms:

\begin{eqnarray}
\frac{\delta \Wb_{\bar{\epsilon}}}{\delta \eb^{a}} &=&
\frac{i}{2}\bar{\epsilon}_{A}\gamma^{5} \gamma_{a} \cex \pb^{A}
-\frac{i}{2}\cex\bar{\epsilon}_{A} \we\gamma^{5} \gamma_{a}
\pb^{A}- i\kappa \frac{\partial \st \Fb }{ \partial \eb^{a}} 
\we \bar{\epsilon}_{A} \ve^A_{\ B} \pb^B \label{weps1}\\
\frac{\delta \Wb_{\bar{\epsilon}}}{\delta \Ab} &=& -i\kappa \ex\left(
\bar{\epsilon}_{A} \ve^A_{\ B} \gamma^{5}\pb^B\right) = -i\kappa \cex\left(
\bar{\epsilon}_{A} \ve^A_{\ B} \gamma^{5}\pb^B\right)\label{weps2}
\end{eqnarray}

Following the identity (\ref{idensi}), these equations have to be
equal respectively to $\frac{\partial}{\partial \ex \eb^{a}}
\left(\bar{\delta} _{1} \fib^{i}\we \frac{\delta \L_{2}}{\delta
\fib^{i}} \right)$ and to $\frac{\partial}{\partial \ex \Ab}
\left(\bar{\delta} _{1} \fib^{i}\we \frac{\delta \L_{2}}{\delta
\fib^{i}} \right)$, where $\fib^{i}$ goes for all the fields (even auxiliary).
Using now (\ref{eqmoe}-\ref{eqmoa}) together with
(\ref{ndet1}-\ref{ndet3}), the pair of equations
(\ref{weps1}-\ref{weps2}) has to be equal to:

\begin{eqnarray}
 (\ref{weps1})&=& -\frac{i}{2}\bar{\delta}_{1} \bar{\pb}_A\we
\gamma ^{5}\gamma _{a} \pb^{A}+ \bar{\delta}_{1} \Ab \we
\frac{\partial \ex \st \Fb}{\partial \ex \eb^{a}}\nonumber\\
&\ & +\frac{\partial}{\partial \ex \eb^{a}}
\left(\bar{\delta}_{1} \w^{cd}
\we \frac{\delta \L_{2}}{\delta
\w^{cd}} + \bar{\delta}_{1} \Fb\we \frac{\delta \L_{2}}{\delta
\Fb}\right)\label{compl1}\\
(\ref{weps2})&=& \bar{\delta} _{1} \eb^{b} \we \frac{\partial
\st\Fb}{\partial \eb^{b}}-i\kappa \bar{\delta}_{1}
\bar{\pb}_A\we \ve^A_{\ B} \gamma ^{5}\pb^B \nonumber\\
&\ & + \frac{\partial}{\partial \ex \Ab} \left(\bar{\delta}_{1} \w^{cd}
\we \frac{\delta \L_{2}}{\delta
\w^{cd}} + \bar{\delta}_{1} \Fb\we \frac{\delta \L_{2}}{\delta\Fb}\right)\label{compl2}
\end{eqnarray}

Using now that $\frac{\partial \ex \st \Fb}{\partial \ex
\eb^{a}}=\frac{\partial \st \Fb}{\partial  \eb^{a}}$ and the explicit
expressions for $\bar{\delta} _{1} \eb^{a}$,
$\bar{\delta}_{1} \bar{\pb}_A$ et $\bar{\delta}_{1} \Ab$ we get from
the previous two equations:

\begin{eqnarray}
\frac{\partial}{\partial \ex \eb^{a}}
\left(\bar{\delta}_{1} \w^{cd}
\we \frac{\delta \L_{2}}{\delta
\w^{cd}} + \bar{\delta}_{1} \Fb\we \frac{\delta \L_{2}}{\delta
\Fb}\right) &=& \frac{i\kappa }{2} \bar{\epsilon }_{A}
\left(\gamma ^{5}\Fb_{a}+ \st \Fb_{a} \right)\we \ve^A_{\
B} \pb^B \nonumber\\
&\ & + \frac{i}{2}\bar{\epsilon}_{A}\gamma^{5} \gamma_{a}
\hat{\cex} \pb^{A} \label{preeg1}\\
\frac{\partial}{\partial \ex \Ab} \left(\bar{\delta}_{1} \w^{cd}
\we \frac{\delta \L_{2}}{\delta
\w^{cd}} + \bar{\delta}_{1} \Fb\we \frac{\delta
\L_{2}}{\delta\Fb}\right) &=& 
-i\kappa \bar{\epsilon }_{A} \ve^A_{\
B}\gamma ^{5} \hat{\cex} \pb^{B} +\bar{\delta}_{1} \eb^{a}\we \st \Fb_{a} \nonumber\\
&\ & - \bar{\delta} _{1} \eb^{b} \we \frac{\partial
\st\Fb}{\partial \eb^{b}}
\label{preeg2}
\end{eqnarray}

Now the unknown quantities in the above equations are
$\bar{\delta}_{1} \w^{cd}$ and $\bar{\delta}_{1} \Fb$. To find them,
we can remark that the pair of equations (\ref{preeg1}-\ref{preeg2})
is a special case of (\ref{rrro}), where now the variable $x^{\alpha}$
goes for $\ex \eb^{a}$ and $\ex \Ab$ and $\omega^{i}$ for $\w^{ab}$ and
$\Fb$. As in the ${\cal{N}}_{4}=1$ case, we see from
(\ref{preeg1}-\ref{preeg2}) that $E_{i}$ and $R_{\alpha}$ are
respectively  linear and independent of
$x^{\alpha}$. We can then choose $\bar{\delta}_{1} \omega ^{i}$ to be
independent of $x^{\alpha}$ (see discussion after equation (\ref{rrro})). In that case, the pair
(\ref{preeg1}-\ref{preeg2}) becomes after some algebra:

\begin{eqnarray}
-\frac{1}{4\kappa ^{2}}\bar{\delta}_{1} \w^{bc} \we \S_{abc}  &=&  \frac{i}{2}\bar{\epsilon}_{A}\gamma^{5} \gamma_{a}
\hat{\cex} \pb^{A}\nonumber\\
&\ & - \frac{i\kappa }{4}
\bar{\epsilon}_{A}\left(F^{bc}-\gamma^{5}\st F^{bc} \right)\ve^A_{\ B}
\pb^{B}\we \S_{abc} \label{contf1}\\
\bar{\delta}_{1}\left(\st \Fb \right) -\bar{\delta}_{1} \eb^{a}\we \st
\Fb_{a} &=& -i\kappa \bar{\epsilon }_{A} \ve^A_{\
B}\gamma ^{5} \hat{\cex} \pb^{B} \label{contf2}
\end{eqnarray}

The last steep is to isolate
$\bar{\delta}_{1} \w^{ab}$ and $\bar{\delta}_{1} \Fb$ from
(\ref{contf1}-\ref{contf2}). The lhs and the
first term of the rhs of (\ref{contf1}) are analogous to the result
(\ref{solw}) of ${\cal{N}}_{4}=1$ supergravity. Then
it can be inverted using the matrix (\ref{defm}), giving three terms
analogous to the Deser-Zumino ones. The second term of the rhs of
(\ref{contf1}) is proportional to $\S_{abc}$ (as the lhs) and thus
does not require any more work:

\begin{eqnarray}
\bar{\delta}_1 \w_{ab} &=& 2 i\kappa^2 \bar{\epsilon} \gamma^5 \left(\gab_{(1)} \st\hat{\psi}_{ab} 
-\frac{1}{2} \eb_a \gamma^c \st\hat{\psi}_{cb}
+ \frac{1}{2} \eb_{b} \gamma^c \st\hat{\psi}_{ca} \right) \nonumber\\
& & +i\kappa^{3}\bar{\epsilon}_{A}\left(F_{ab}-\gamma^{5}\st F_{ab}
\right)\ve^A_{\ B} \pb^{B}\label{desbis}
\end{eqnarray}
where $ \st\hat{\psi}_{\mu\nu} =\frac{1}{2} \ve_{\mu\nu\rho\sigma} \hat{D}^{\rho}
\psi^{\sigma}$ and $\st F_{\mu \nu }:=\frac{1}{2}\ve_{\mu \nu \rho
\sigma } F^{\rho \sigma }$.

To invert the second equation (\ref{contf2}), we use the
identity: $\delta \left(\st \Fb \right)-\delta
\eb^{a}\we  \st \Fb_{a}=\st\left(\delta \Fb -\delta \eb^{a}\we
\Fb_{a}\right)$, which can be proved straightforwardly, see also
\cite{SiT}. Then equation (\ref{contf2}) implies that:

\begin{equation}\label{finalfff}
\bar{\delta}_{1}\Fb = i\kappa \bar{\epsilon }_{A} \ve^A_{\ B}\gamma
^{5}\st \left( \hat{\cex} \pb^{B} \right) +
\bar{\delta}_{1}\eb^{a}\we \Fb_{a}
\end{equation}

Thus, the $1^{st}$ order ${\cal{N}}_{4}=2$ Lagrangian (\ref{lagsugra})
is invariant under the supersymmetry transformation laws given by
(\ref{ndet1}-\ref{ndet3}), (\ref{desbis}) and (\ref{finalfff}).

It is interesting to note that using the new method it is possible to
find a nice formula for the supersymmetry transformation laws, in the
sense that they depend on the independent auxiliary fields (namely
$\w^{ab}$ and $\Fb$) and not on the induced auxiliary fields (namely
$\w^{ab}_{2}$ and $\Fb_{2}$ or $\ex \eb^{a}$ and $\ex \Ab$). 
This aesthetical criterion was the motivation for looking for a new method
in section \ref{newm}, and not simply stop with the general results
(\ref{defd1}), (\ref{nouvva}) and (\ref{trancoo}). In the next
example, namely the ${\cal{N}}_{11}=1$ supergravity, we will prove
that such an ``aesthetical'' result does not exist.

\section{$1^{st}$ order formalism for ${\cal{N}}_{11}=1$ supergravity}

\subsection{The $1^{st}$ order Lagrangian }\label{flaghh}

The five basic fields of the $1^{st}$ order eleven dimensional
supergravity are the  one-form vierbein $\eb^a$, the so(1,10) one-form
(spin) connection $\w^a_{\ b}$, the Rarita-Schwinger one-form $\pb$
and the three-form tensor $\Ab=\frac{1}{3!} A_{\m\n\r} \ex x^\m \we
\ex x^\n \we \ex x^\r$ and its associated field strength
$\Fb=\frac{1}{4!} F_{\m\n\r\sigma } \ex x^\m \we
\ex x^\n \we \ex x^\r\we \ex \sigma $. The eleven-form Lagrangian can
be rewritten as\footnote{Note that (\ref{lag11}) (whose gravitational
part was first proposed in \cite{CFGPN}) differs slightly from
the original formulation 
of eleven dimensional supergravity of \c{CJS}: In fact (\ref{lag11})
is just the standard Lagrangian reexpressed in terms of the new connection
$\w^a_{\ b}$. This connection, as an independent variable, is {\it
not} the independent 
connection of \c{CJS} but is shifted in such a way that its induced
value
is what was called in that paper
$\hat{\w}^a_{\ b}$. In this form, the equations of motion of the
connection (\ref{eqmow11}) are automatically supercovariant, in the
sense that their supersymmetry variation
(\ref{tranp11}-\ref{trana11}) does not contain the
derivative of the gauge parameter. Finally, $\L_{11}$ of (\ref{lag11})
is a complete $1^{st}$ order formulation, with the four form also
treated as an independent field. We should emphasize that the use of a
supercovariant connection is a technical help not a fundamental
requirement in this work. }:

\begin{eqnarray}
\L_{11} &=& -\frac{1}{4\k^2} \R^{ab} \we \S_{ab} + \frac{i}{2}
\bar{\pb}\we \gab_{(8)} \we \cex \pb +\frac{i}{8} \left( \X^a -
\frac{i \k^2}{4} \bar{\pb} \g^a \we \pb \right) \we \eb_a \we
\bar{\pb} \we \gab_{(6)} \we \pb\nonumber\\
&\ & - \frac{1}{2} \Fb \we \st \Fb +\left(\st \Fb +\bb \right)\we
\left(\ex \Ab - \ab \right) + \frac{1}{2}\ab\we \bb -
\frac{\kappa }{3} \Ab\we\ex  \Ab\we\ex\Ab \label{lag11}
\end{eqnarray}
with the definitions:
\begin{equation}\label{defabb}
\ab:=\frac{i\kappa }{4} \bar{\pb}\we \gab_{(2)} \we \pb
\hskip.9cm
\bb:=\frac{i\kappa}{4}\bar{\pb}\we\gab_{(5)} \we \pb
\end{equation}

We also defined $\gab_{(n)} := \frac{1}{n!} \gab_{(1)}\we \gab_{(1)} \we \ldots \we \gab_{(1)}$ ($n$ times). The
notations and conventions are the same as in the previous subsections,
generalized straightforwardly to eleven dimensions\footnote{Namely
$\g^0 \g^1 \ldots \g^{10}=1=\ve_{01\ldots10}$ and $\eta^{ab}=\{
-,+,\ldots,+\}$, see also appendix A.}. 
The equations of
motion for (\ref{lag11}) are given by:

\begin{eqnarray}
\frac{\d \L_{11}}{\d \eb^a} &=& -\frac{1}{4 \k^2} \R^{bc} \we
\S_{bca} -\frac{i}{4} \bar{\pb} \we (2 \gab_{(7)a}+\eb_{a}\we \gab_{(6)}) \we \cex \pb\nonumber\\
& & +\frac{i}{4} \X^b\we \bar{\pb} \we \left(  \gab_{(6)} \eta_{ab} + \gab_{(5) (a}
\we \eb_{b)} \right) \we \pb + \t_a = 0 \label{eqmoe11}\\
\frac{\d \L_{11}}{\d \bar{\pb}} &=& i  \gab_{(8)} \we \hat{\cex} \pb- \frac{i\kappa }{2}\gab_{(5)}\we \pb\we \left(\Fb-\ex \Ab + \ab
\right) \nonumber\\
& & +\frac{i}{4}  \left( \X^a -  \frac{i \k^2}{2} \bar{\pb} \g^a\we
\pb \right) \we \left( 2 \gab_{(7)a} -\gab_{(6)} \we \eb_a \right) \we
\pb = 0 \label{eqmop11}\\
\frac{\d \L_{11}}{\d \Ab} &=& \ex \left(\st \Fb+\bb\right) - \k
\ex \Ab  \we \ex \Ab = 0  \label{eqmoa11}\\
\frac{\d \L_{11}}{\d \w^{ab}} &=& -\frac{1}{4 \k^2} \left( \X^c -
\frac{i \k^2}{2} \bar{\pb} \g^c\we \pb \right) \we \S_{abc} = 0 \label{eqmow11}\\
\frac{\delta \L_{11}}{\delta \Fb} &=& \st\left(\ex \Ab - \ab - \Fb
\right) = 0 \label{eqmof11}
\end{eqnarray}
where we also used the shorthand notation defined by (\ref{defintp}).

Some additional comments are needed:

\begin{enumerate}
\item [] $\bullet$ The Lagrangian (\ref{lag11}) can be rewritten (up
to a surface term) in
the following suggestive form:
\begin{eqnarray}
\L_{11} &=& -\frac{1}{4\k^2} \left[\frac{1}{2} \dia\w_{a}\we \w^{a} + (\dia
\w_{a} +\beb_{a})\we \left(\ex \eb^{a}-\alb^{a} \right)
+\frac{1}{2} \alb^{a}\we \beb_{a}  \right] + \frac{i}{2}
\bar{\pb}\we \gab_{(8)} \we \ex \pb \nonumber\\
&\ & - \frac{1}{2} \Fb \we \st \Fb +\left(\st \Fb +\bb \right)\we
\left(\ex \Ab - \ab \right) + \frac{1}{2}\ab\we \bb -
\frac{\kappa }{3} \Ab\we\ex  \Ab\we\ex\Ab \label{lag12}
\end{eqnarray}
where we defined $\w^{a}:=\w^{a}_{\ b}\we \eb^{b}$,
$\dia\w_{a}:=\w^{bc}\we \S_{abc}$ and 
\begin{equation}\label{defalp}
\alb^{a}:=\frac{i\kappa^{2}}{2} \bar{\pb}\gamma^{a}\we \pb
\hskip.4in \beb_{a}:=-\frac{i\kappa^{2}}{2} \eb_{a}\we \bar{\pb}\we \gab_{(6)}\we \pb
\end{equation}

Note that the $\dia$ operation is up to a trace, a Hodge-duality
operator: from the definitions, $\dia\w_{a}=\st (\w_{a}
+\eb^{b}\we \i_{a} \w_{b}-2 \eb_{a}\we \i_{b} \w^{b})$.
\item [] $\bullet$ The energy-momentum tensor of (\ref{eqmoe11}) is given by $\t_{a}=\frac{\partial}{\partial \eb^{a}}\left(  \st \Fb + \bb \right)  \we \ex
\Ab+ \frac{\partial}{\partial \eb^{a}}\left(-\st \Fb\we
\ab-\frac{1}{2}\Fb\we \st \Fb 
+\frac{\kappa ^{2}}{16}\bar{\pb}\we \gab_{(1)}\we\pb\we\bar{\pb}
\we \gab_{(6)}\we \pb\right)$.
\item [] $\bullet$ The ``supercovariant-hatted'' derivative was
derived in \c{CJS} and is given by: $\hat{\cex}
\pb:= \cex  \pb - \frac{\k}{12^2}
\left(\g_{abcde}+ 8 \g_{bcd}\eta_{ea}\right) \eb^{a} F^{bcde} \we
\pb$. The equation of motion (\ref{eqmop11}) can be computed only
after making use of
\begin{enumerate}
\item [] a Fierz rearrangement: 

$4\gab_{(7)a}\we \pb\we (
\bar{\pb}\g^a\we \pb)+ 
\gab_{(1)}\we \pb\we (\bar{\pb}\we\gab_{(6)}\we\pb)-
\gab_{(6)}\we \pb\we (\bar{\pb}\we\gab_{(1)}\we\pb)+
\gab_{(5)}\we \pb\we (\bar{\pb}\we\gab_{(2)}\we\pb)-
\gab_{(2)}\we \pb\we (\bar{\pb}\we\gab_{(5)}\we\pb)=0$
\item [] a supercovariant derivative identity: 

$i\gab_{8}\we \hat{\cex} \pb = i\gab_{8}\we\cex\pb
+\frac{i\kappa }{2}\left(\gab_{(5)}\we \Fb+\gab_{(2)}\we \st \Fb
\right)\we\pb$.
\end{enumerate}

As usual, the last two terms in the rhs of (\ref{eqmop11}) vanish if
the equations of motion of the auxiliary fields $\w^{ab}$ and $\Fb$ are used.
\item [] $\bullet$ We fixed the action in such a way that the 
induced auxiliary fields $\w^{ab}_{2}$ and $\Fb_{2}$
(computed as the only solutions of
(\ref{eqmow11}-\ref{eqmof11})) are supercovariants, in the sense that
their supersymmetry transformation laws (induced by
(\ref{tranp11}-\ref{trana11})) do not contain the derivatives of the
gauge parameter $\bar{\epsilon}$. Note however that we paid a price
for this, namely we changed the quartic fermionic terms which seem 
unavoidable beyond 4 dimensions for pure supergravities. In the
previous example the graviphoton did also cause such a complication.
\end{enumerate}

\subsection{The supersymmetry transformation law}\label{ghjk}

The
Lagrangian (\ref{lag11}) is invariant in a $2^{nd}$ order sense (ie
after using freely the auxiliary fields equations of motion)  under
the following 
supersymmetry transformation laws:

\begin{eqnarray}
\d_2 \bar{\pb} &=&  \hat{\cex}_{2} \bar{\e}:= \cex_{2} \bar{\e} +
\frac{\k}{12^2} \bar{\e}\left(\g_{abcde}- 8 \g_{bcd}\eta_{ea}\right)\eb^{a}
F^{bcde}_{2} \label{tranp11}\\
\d_2 \eb^a &=& i \k^2 \bar{\e} \g^a \pb\label{trane11}\\
\d_2 \Ab &=& \frac{i\k}{2}  \bar{\e} \gab_{(2)} \we \pb\label{trana11}
\end{eqnarray}

As before, a $1^{st}$ order formulation of eleven dimensional
supergravity is given by equations (\ref{tranp11}-\ref{trana11})
(following  (\ref{defd1})), and by  equations  (\ref{nouvva}) and
(\ref{trancoo}) for the auxiliary fields. In a more explicit way,
these are:

\begin{eqnarray}\label{autran11}
\delta_{1}\w_{ab} &=&-i\kappa^{2}
\bar{\epsilon}\left(\gamma_{c}\hat{\psi}_{2ab}+\gamma_{a}\hat{\psi}_{2cb}-\gamma_{b}\hat{\psi}_{2ca}
\right)\eb^{c}+\frac{1}{2} \left(T_{a,bc}-T_{c,ab}+T_{b,ca}\right) \eb^{c}\nonumber\\
& & + \frac{i\kappa^{3}}{72}\bar{\epsilon}\left(\gamma_{abc_{1}\dots 
c_{4}}+24 \eta_{ac_{1}}\eta_{bc_{2}}\gamma_{c_{3}c_{4}}
\right)F_{2}^{c_{1}\dots c_{4}} \pb 
\end{eqnarray}
where $\hat{\psi}_{2ab}:=\hat{D}_{2[a} \psi_{b]}$ and
$\hat{\cex}_{2}$ is the $2^{nd}$ order supercovariant derivative
constructed with the induced auxiliary fields, namely 
$\w^{ab}_{2}$ and $\Fb_{2}$, see (\ref{tranp11}). Moreover, the tensor
$T_{a,bc}$ (which 
vanishes modulo the equation of motion of the connection) was defined
just after equation (\ref{nouvva}).

The transformation law for the field strength can also be computed
following equation (\ref{trancoo}). The final result is\footnote{To
find the result (\ref{ffftr}), we need the following
eleven-dimensional Fierz identity: $\gamma^{a}\pb\we \bar{\pb}\we
\gab_{(1)a}\we \pb + \gab_{(1)a}\we\pb \we \bar{\pb} \gamma^{a}\we
\pb=0$, which can be proved using the general formula (\ref{fierz})
with $\pb_{1}=\pb$, $\bar{\pb}_{2}=\bar{\pb}$ and
$\pb_{3}=\gab_{(1)a}\we \pb$ for the first term and  $\pb_{1}=\pb$, $\bar{\pb}_{2}=\bar{\pb}$ and
$\pb_{3}=\gamma_{a} \pb$ for the second one.}:

\begin{equation}\label{ffftr}
\delta_{1}\Fb =\frac{i\kappa }{2}\bar{\epsilon}\gab_{(2)}\we
\hat{\cex}_{2}\pb + \delta_{2}\eb^{a}\we \i_{a}\Fb
-\frac{\delta_{2} e}{2e} \Delta \Fb
\end{equation}
with $e$ the determinant of the vielbein.

A $1^{st}$ order formulation for the gravitational part of
eleven-dimensional supergravity was first derived in \cite{CFGPN}, it
did not treat however the four form field strength as an independent
field. 
It
differs from (\ref{tranp11}-\ref{autran11}) by some trivial gauge
transformation of the type (\ref{twist}) (to see how to proceed, it is
enough to follow the method given in section \ref{thsussy}). We could
also add another trivial gauge transformation to replace the $\Fb_{2}$
by $\Fb$ in (\ref{tranp11}), and then $\hat{\cex}_{2}$ by
$\hat{\cex}$. The net result would be to add a contribution to
(\ref{ffftr}) proportional to the equations of motion of the
gravitino. 

It is not obvious at all that the final expressions for
$\delta_{1}\w^{ab}$ and $\delta_{1} \Fb$ would be simpler or ``more
geometrical'' after adding these trivial gauge transformations. The
question we would like to answer is thus to know whether there exists some
appropriate trivial gauge transformation, which, added to
(\ref{tranp11}-\ref{ffftr}), gives as ``nice'' a result as in the 
 ${\cal{N}}_{4}=2$ supergravity case(\ref{desbis}-\ref{finalfff}).

By ``nice'' we mean the following: we know (from the new method of
section \ref{newm}) that 
there exist transformation laws
$\bar{\delta}_{1}\w^{ab}$ and $\bar{\delta}_{1}\Fb$ which,
together with 
\begin{eqnarray}
\bar{\delta}_{1} \bar{\pb} &:=& \hat{\cex} \bar{\epsilon}\label{ndet11}\\
\bar{\delta}_{1} \eb^{a} &:=& \delta_{2}\eb^{a}\label{ndet21}\\
\bar{\delta}_{1} \Ab &:=& \delta_{2}\Ab\label{ndet31}
\end{eqnarray}
leave the Lagrangian (\ref{lag11}) invariant.

A ``nice'' formula would be one where these $\bar{\delta}_{1}\w^{ab}$
and $\bar{\delta}_{1}\Fb$ would be independent of $\ex \eb ^{a}$ (ie
independent of $\Delta \w^{ab}$) and of $\ex \Ab$ (ie independent of
$\Delta \Fb$) respectively. We will end this section by giving strong
arguments supporting the idea that such a ``nice'' formula does not exist.

As in previous examples, the $(D-1)$-form defined by (\ref{defw})
is given by formula (\ref{wsusy})\footnote{Remember that if the induced auxiliary fields are
supercovariants, so are the auxiliary fields themselves, and then, only
the gravitino transformation law contains
$\ex\bar{\epsilon}$-terms.}. Now, using the gravitino's equation of
motion (\ref{eqmop11}), we can compute the eleven dimensional version
of the pair of equations (\ref{weps1}-\ref{weps2}):

\begin{eqnarray}
\frac{\delta \Wb_{\bar{\epsilon}}}{\delta \eb^{a}} &=&
\frac{i}{4}\bar{\epsilon}\left(2\gab_{(7)a}+\eb_{a}\we \gab_{(6)}
\right)\we \cex \pb + \frac{i}{4}\cex \bar{\epsilon}\left(2\gab_{(7)a}-\eb_{a}\we \gab_{(6)}
\right)\we \pb\nonumber\\
& & -\frac{i}{2} \X^{b}\bar{\epsilon}\left(\eta_{ab}\gab_{(6)}
+\gab_{(5) (a}\we\eb_{b)} \right) + \frac{i\kappa }{2}
\bar{\epsilon} \gab_{(4) a} \we  \pb \we \ex \Ab
+ \mbox{\it more} \label{zzzwww1}\\
\frac{\delta \Wb_{\bar{\epsilon}}}{\delta \Ab} &=& \frac{i\kappa }{2}
\cex\left(\bar{\epsilon} \gab_{(5)}\we \pb \right)\label{z2zzwww}
\end{eqnarray}
where {\it ``more''} goes for terms proportional to
$\bar{\epsilon}\pb^{3}$ and $\bar{\epsilon}\pb \Fb$ and
 useless for our purpose. 

Using the identity (\ref{idensi}), the equations
(\ref{zzzwww1}-\ref{z2zzwww}) are respectively equal to:

\begin{eqnarray}
  \frac{i}{4}\bar{\delta}_{1} \bar{\pb} \we\left(2
\gab_{(7)a}-\eb_{a}\we \gab_{(6)} \right) \pb +\mbox{\it more}+\frac{\partial}{\partial \ex \eb^{a}}
\left(\bar{\delta}_{1} \w^{cd}
\we \frac{\delta \L_{2}}{\delta
\w^{cd}} + \bar{\delta}_{1} \Fb\we \frac{\delta \L_{2}}{\delta
\Fb}\right)\label{czompl1}\\
 \frac{i\kappa}{2} \bar{\delta}_{1}
\bar{\pb}\we  \gab_{(5)}\we \pb -2 \kappa \bar{\delta}_{1}\Ab
\we \ex \Ab +\mbox{\it more}+
\frac{\partial}{\partial \ex \Ab} \left(\bar{\delta}_{1} \w^{cd}
\we \frac{\delta \L_{2}}{\delta
\w^{cd}} + \bar{\delta}_{1} \Fb\we \frac{\delta \L_{2}}{\delta\Fb}\right)\label{czompl2}
\end{eqnarray}
and {\it ``more''} has the same meaning as before.

The above two equalities are a special case of the general equation
(\ref{rrro}), where $x^{\alpha}$ goes for $\ex \eb ^{a}$ and $\ex \Ab$
and $\omega^{i}$ for $\w^{ab}$ and $\Fb$. As before, the auxiliary field
equations of motion (\ref{eqmow11}-\ref{eqmof11}) $E_{i}$ are linear in
$x^{\alpha}$ (namely in $\ex \eb ^{a}$ and $\ex \Ab$), ie
$E_{i}=E_{i\alpha}x^{\alpha}+E_{i0}$. The very big difference with the
four dimensional supergravities is that now $R_{\alpha}$ is also {\it linear}
in (and {\it not} independent of)  $x^{\alpha}$ (by rearranging of equations
(\ref{zzzwww1}-\ref{czompl2})), $R_{\alpha}=R_{\alpha \beta}
x^{\beta}+R_{\alpha 0}$. That means that $\bar{\delta}_{1}\omega^{i}$
cannot be independent of $x^{\alpha}$. Instead, we can choose it to
be linear in this variable $x^{\alpha}$. The practical reasons for
that annoying feature are the gammalogy beyond four dimensions and
also the presence of the Chern-Simons term.

This  means in particular that $\bar{\delta}_{1}\w^{ab}$
are linear in $\ex \eb^{a}$, and then in $\w^{ab}_{2}$ (it is always
possible to rewrite the derivative of the vierbein in term of the induced
connection). So it will be impossible to rewrite $\bar{\delta}_{1}\w^{ab}$
in a ``nice'' way (nicer than the straightforward result given by
(\ref{tranp11}-\ref{ffftr})), in the sense that
only the independent auxiliary field appears in the transformation
laws and not the induced one. The same is true for $\Fb$ and $\ex
\Ab$. 

This problem seems to appear as soon as we go beyond 4 dimensions.
The  reason for that remains
however unclear for us. 
On the other hand this may have some importance for the construction
of superspace versions of these theories.
 A first attempt at a deeper
understanding should address the 5 dimensional case where auxiliary
fields are in principle known.

For completeness we should also mention related works by Bars,
MacDowell, Higuchi and Kallosh \cite{Ka} where the (partially)
first order formalism of \cite{CFGPN} is combined with an
identification of the three form gauge field with supertorsion so as
to lead to more equivalent forms of classical 11 dimensional supergravity.
Recently a preprint \cite{NVN} appeared where some first order results
for the three form sector of 11d supergravity are listed, they should
be obtainable from our formulas by elimination of the Lorentz connection.

\bigskip
{\bf Acknowledgements.}

\bigskip
We benefited from conversations or correspondance with E. Cremmer,
M. Henneaux and  W. Siegel.

We are grateful to P. van Nieuwenhuizen and P. West for historical
comments.

\appendix

\bigskip

\section{Conventions and useful formulas }

The conventions are the following for a $D$-dimensional spacetime
dimension (see \cite{Wes} for a more complete discussion):

$\eta_{ab}=\{-,+,\ldots,+\}$,
$\ve_{01\ldots(D-1)}=1$, $\g_a$ ($a=0,\ldots, D-1$) are $D$ gamma matrices of
dimension $2^{[\frac{D}{2}]}$ which satisfy the Clifford algebra $\left\{
\g_a, \g_b \right\} = 2 \eta_{ab}$. Here, $[r]$ is the integer part of $r$
(i.e. [4.5]=4). If $D$ is even (odd), $\g^{D+1} (\1) := \g^0 \ldots
\g^{D-1}$, $\left(\g^{D+1}\right)^2= (-)^{\frac{D (D-1)}{2}+1} \1$. 

Some definitions and useful $\g$ matrix product formulas are:

\B \g_{a_1 \ldots a_n}:=\g_{[a_1}\ldots \g_{a_n]} \E

we define by $\gab_{(n)}$ the $(n)$-form:

\B \gab_{(n)}:= \frac{\g_{a_1\ldots a_n}}{n!}\eb^{a_1}\we \ldots\we \eb^{a_n} \E

and also

\begin{equation}\label{defintp}
\gab_{(n) a_1\ldots a_m} := \i_{a_1} \ldots \i_{a_m} \gab_{(n+m)} :=\g_{a_1\ldots a_m b_1\ldots b_n}
\frac{\eb^{b_1}\we \ldots\we
\eb^{b_n}}{n!}
\end{equation}

for any integers $m$ and $n$, $\i_a$... mean contraction with the
relevant frame vector.

Now, from the standard decomposition formula (complete
antisymmetrisation in $a_m\ldots a_1$ and $b_1\ldots
b_n$ is understood in the above formulas):

\B \g_{a_m\ldots a_1} \g_{b_1\ldots b_n} = \sum_{i=0}^{min(m,n)}
i!
\left(\begin{array}{c} m \\ i \end{array} \right)
\left(\begin{array}{c} n \\ i \end{array} \right)
\eta_{a_1 b_1} \ldots \eta_{a_i
b_i} \g_{a_m \ldots a_{i+1} b_{i+1} \ldots b_n} \E

it is easy to derive:

\begin{eqnarray}
\gab_{(m)}\we \gab_{(n)} &=&  
\left(\begin{array}{c} m+n \\ n \end{array} \right)\gab_{(m+n)}\\
\g_{a_m\ldots a_1} \gab_{(n)} &=& \sum_{i=0}^{min(m,n)} 
\left(\begin{array}{c} m \\ i \end{array} \right)
\eb_{a_1}\we \ldots \we \eb_{a_i}\we \gab_{(n-i) a_m \ldots a_{i+1}} \\
 (-)^{m.n} \gab_{(n)} \g_{a_m\ldots a_1} &=& \sum_{i=0}^{min(m,n)}
\left(\begin{array}{c} m \\ i \end{array} \right)
 (-)^i \eb_{a_1}\we \ldots \we \eb_{a_i}\we
\gab_{(n-i) a_m \ldots a_{i+1}}
\end{eqnarray}

where $\left(\begin{array}{c} m \\ i \end{array}
\right)$ stands for $\frac{m!}{(m-i)! i!}$.

The general Fierz formula in any spacetime dimension $D$ is given by:

\begin{equation}\label{fierz}
\pb_1\we \left( \bar{\pb}_2\we \pb_3 \right) =
\frac{(-)^{1+p_1\pt p_2+p_2\pt p_3+p_3\pt p_1}}{2^{[\frac{D+1}{2}]}}
\sum_{n=0}^D \frac{1}{n!}\g^{a_n\ldots a_1} \pb_3\we\left(
\bar{\pb}_2\we \g_{a_1 \ldots a_n} \pb_1 \right)
\end{equation}

Where $\pb_i$ is a $(p_i)$-form anticommuting spinor ($i=1, \ldots,
3$). Note that the summation in (\ref{fierz}) contains ($D+1$)
terms. In the case where $D$ is odd, half of them can be eliminated by
a Hodge-duality transformation (recall that in that case
$\g^{D+1}\sim \1$). Then the summation goes only up to $\frac{D-1}{2}$ and
a factor of 2 automatically  cancels out in the denominator of (\ref{fierz}).  

The Majorana (form)-spinors (which exist in $D=4$ and $D=11$) satisfy the symmetry property:

\B \bar{\pb}_1 \we \g^{a_1 \ldots a_n} \pb_2 =
(-)^{p_1\pt p_2+\frac{n(n+1)}{2}}  \bar{\pb}_2 \we \g^{a_1 \ldots a_n}
\pb_1 \lb{ants}\E

\end{document}